\documentclass[aps,twocolumn,amsmath,amssymb,showpacs,prb,superscriptaddress,unsortedaddress]{revtex4}
\usepackage{epsf}
\usepackage{graphicx}

\begin{document}

\title{Controlling the carrier concentration of the high temperature superconductor Bi$_2$Sr$_2$CaCu$_2$O$_{8+\delta}$ in Angle Resolved Photoemission Spectroscopy (ARPES) experiments}

\author{A. D. Palczewski}
\affiliation{Ames Laboratory and Department of Physics and Astronomy, Iowa State University, Ames, IA 50011, USA}

\author{T. Kondo}
\affiliation{Ames Laboratory and Department of Physics and Astronomy, Iowa State University, Ames, IA 50011, USA}

\author{J. S. Wen}
\affiliation{Condensed Matter Physics and Materials Science Department, Brookhaven National Laboratory, Upton, New York 11973, USA }

\author{G. Z. J. Xu}
\affiliation{Condensed Matter Physics and Materials Science Department, Brookhaven National Laboratory, Upton, New York 11973, USA }

\author{G. Gu}
\affiliation{Condensed Matter Physics and Materials Science Department, Brookhaven National Laboratory, Upton, New York 11973, USA }

\author{A. Kaminski}
\affiliation{Ames Laboratory and Department of Physics and Astronomy, Iowa State University, Ames, IA 50011, USA}

\date{\today}

\begin{abstract}
We study the variation of the electronic properties at the surface of a high temperature superconductor as a function of vacuum conditions in angle resolved photoemission spectroscopy (ARPES) experiments. Normally, under less than ideal vacuum conditions the carrier concentration of Bi$_2$Sr$_2$CaCu$_2$O$_{8+\delta}$ (Bi2212) increases with time due to the absorption of oxygen from CO$_2$  and CO molecules that are prime contaminants present in ultra high vacuum (UHV) systems. We find that in a high quality vacuum environment at low temperatures, the surface of Bi2212 is quite stable (the carrier concentration remains constant), however at elevated temperatures the carrier concentration decreases due to the loss of oxygen atoms from the Bi-O layer. These two effects can be used to control the carrier concentration \textit{in-situ}. Our finding opens the possibility of studying the electronic properties of the cuprates as a function of doping across the phase diagram on the \textit{same} piece of sample (i.e. with the same impurities and defects). We envision that this method could be utilized in other surface sensitive techniques such as scanning tunneling microscopy/spectroscopy.
\end{abstract}


\maketitle

\section{Introduction}

Surface techniques have played an important role in understanding the properties of the high temperature superconductors. They have revealed a number of fascinating phenomena such as the direct observation of the superconducting gap\cite{OLSON} and its anisotropy\cite{SHENSC,HONGSC}, confirmation of the d-wave symmetry of the order parameter, direct observation of the pseudogap and its anisotropy\cite{HONGPG, LOESERPG,MIKEPG}, discovery of spatial inhomogeneities\cite{DAVIS,YAZDANI},unusual spatial ordering,\cite{DAVISCHECKER} nodal quasiparticles\cite{KAMINSKIQP}, renormalization effects\cite{VALLA,BOGDANOV,KAMINSKIKINK} and many others\cite{SHENREVIEW,JCREVIEW}. The success of these techniques rely on the fact that the layers in some cuprates are very weakly bonded via the Van der Waals interaction. In such cases the bulk properties and surface properties are essentially identical, since there is no charge exchange between the layers. The samples in such cases can be thought of as a stack of very weakly electrically coupled 2-dimensional conducting surfaces rather than a 3-dimentional object. Two of the most commonly studied materials with this property are Bi$_2$Sr$_2$CaCu$_2$O$_{8+\delta}$ (Bi2212) and Bi$_2$Sr$_2$CuO$_{6+\delta}$ (Bi2201). There is however one important aspect that needs to be carefully considered, namely the stability of the cleaved samples under ultra high vacuum (UHV) conditions. UHV is a rather broad term and refers to pressures lower than 1$\times$10$^{-9}$ Torr. Quite often such conditions are not sufficient to guarantee the stability of the surface, particularly in the case of non-stoichiometric materials such as the cuprates. These problems were recognized early on\cite{SHEN1}, and subsequent measurements revealed significant changes in the electronic properties as a function of time after cleaving. This issue was not carefully examined following these first measurements, and it is likely an important source of data discrepancies among the various groups \cite{SHENREVIEW,JCREVIEW}.

Here we present a systematic study of the electronic properties of Bi2212 as a function of vacuum conditions.  We demonstrate that under poor vacuum conditions increased carrier concentration arises due to the breakup of CO and CO$_2$ molecules by exposure to vacuum ultra-violet (VUV) photons and the subsequent adsorption of oxygen into the BiO layers. We show that with a UHV leak a sample can increase it carrier concentration just by sitting in the vaccum.  This observation confirms that bilayer splitting is only observed in over-doped Bi2212.  When the partial pressure of active gases is kept at low levels, the lifetime of cleaved surface of Bi2212 can be as long as a few weeks at low temperatures (T$<$150K). At elevated temperatures (T$>$200K) the sample surface loses oxygen, which results in the reduction of carrier concentration. This second effect is most likely responsible for the recently reported non-monotonic temperature dependence of the pseudogap\cite{A. A. Kordyuk 2008}, where at elevated temperatures the sample surface becomes underdoped and therefore develops a pseudogap. We demonstrate that these two effects (\textit{in-situ} absorption and desorption of oxygen) can be utilized to control the carrier concentration of the sample surface. This approach enables one to study the intrinsic electronic properties (i.e. without changing the impurities and defects) of the cuprates across the phase diagram.

\section{Experimental Details}

The ARPES data was acquired using a laboratory-based Scienta 2002 electron analyzer and high intensity Gammadata UV4050 UV source  with custom designed optics. The photocurrent at the sample was approximately 1 $\mu A$, which corresponds to roughly 10$^{13}$ photons/sec at 0.05\% of the bandwidth. The energy resolution was set at 10 meV and momentum resolution at 0.12$^{\circ}$ and 0.5$^{\circ}$ along a direction parallel and perpendicular to the analyzer slits, respectively. Samples were mounted on a variable temperature cryostat (10-300K) cooled by a closed cycle refrigerator. The precision of the sample positioning stage was 1$\mu m$. The partial pressure of the active gases was at the detection limit of the Residual Gas Analyzer (RGA) and the pressure of hydrogen was below 3$\times$10$^{-11}$ Torr. Excellent vacuum conditions were achieved by strict adherence to good vacuum practices, use of UHV compatible materials and a cumulative bake-out time of the system in excess of 6 months. The typical lifetime of the optimally doped Bi2212 surfaces was greater than two weeks after cleaving, defined as less than 5\% change of the superconducting gap (2 meV) at 40K.  The core-level spectra was acquired on the Hermon beam-line at the Synchrotron Radiation Center using a Scienta 2002 end-station. The photon energy was set at 500 eV and energy resolution at 200 meV.

\section{Increasing carrier concentration}

It has been known for some time that aging (increased surface doping) in the cuprates is caused by less than ideal UHV conditions\cite{SHEN1}.  Aging is usually detected by measuring the superconducting gap (the energy gap as defined by the difference between the peak position of a Bi2212 spectrum and the chemical potential measured by a polycrystalline gold sample) as a function of time.  If the gap shifts to a lower binding energy the sample has aged.\cite{H. Ding 1997,P. Schwaller 2000}.  Fig. 1 (a)-(b) shows an example of this where a freshly cleaved Bi2212 single crystal was scanned in a relatively poor vacuum to see how the spectrum changed over time. A shift to lower binding energy as well as a peak suppression was detected showing the sample was aging.  In Fig. 1 (c) the size of the superconducting gap is shown as a function of time. Only when there were VUV photons on the sample did the sample age. While there we no VUV photons on the sample, from 5 hour to 21 hours, the aging stopped. If the non-VUV scanning time is taken out Fig. 1 (d), the magnitude of the gap shows an exponential decay (blue line). While it is know that surface aging of Bi2212 happens in a poor UHV system, only when there were VUV photons on the sample does did sample actually age, signaling that aging is directly related to having VUV photons on the sample, not just the vacuum conditions. 
\begin{figure}
\includegraphics[width=3.6in]{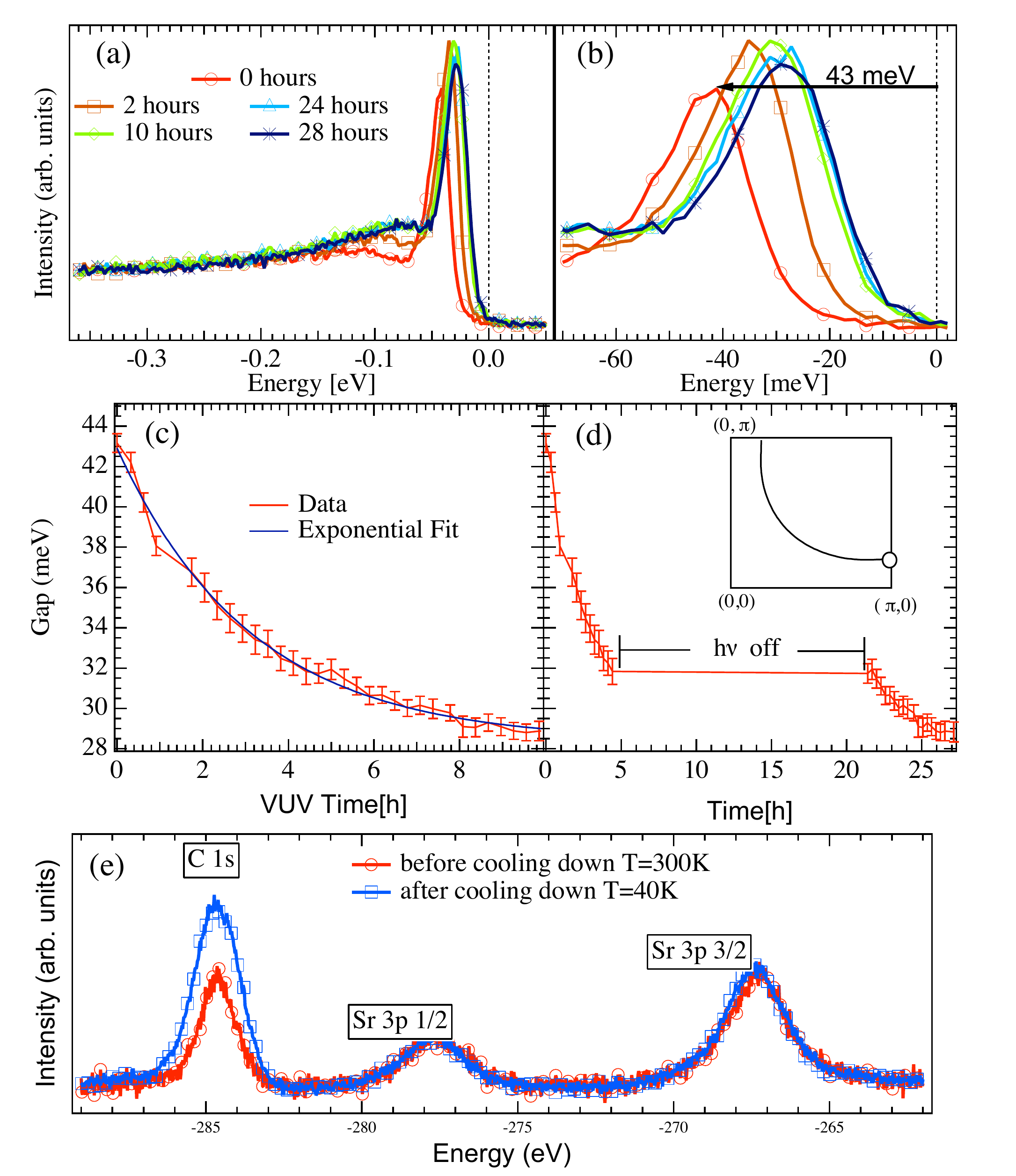}
\caption{Color online) ARPES specta of Bi2212 taken under poor vacuum conditions (a) sample EDC (energy distribution curves) taken at the anti-node where the band crosses the Fermi energy at 5 different times, (b) narrow view of (a), (c) time evolution of Bi2212's superconducting gap as a function of tine, (d) the time evolution of Bi2212's superconducting gap under VUV photons (red) fitted with an exponential decay (blue), the temperature for (a)-(d) was set to 20K,  (e) C 1s, Sr 3P 1/2, Sr 3p 3/2 core level data from Bi2212 showing carbon deposits some time after cleaving and after while cooling.}
\label{Fig. 1}
\end{figure}

\begin{figure}
\includegraphics[width=3.6in]{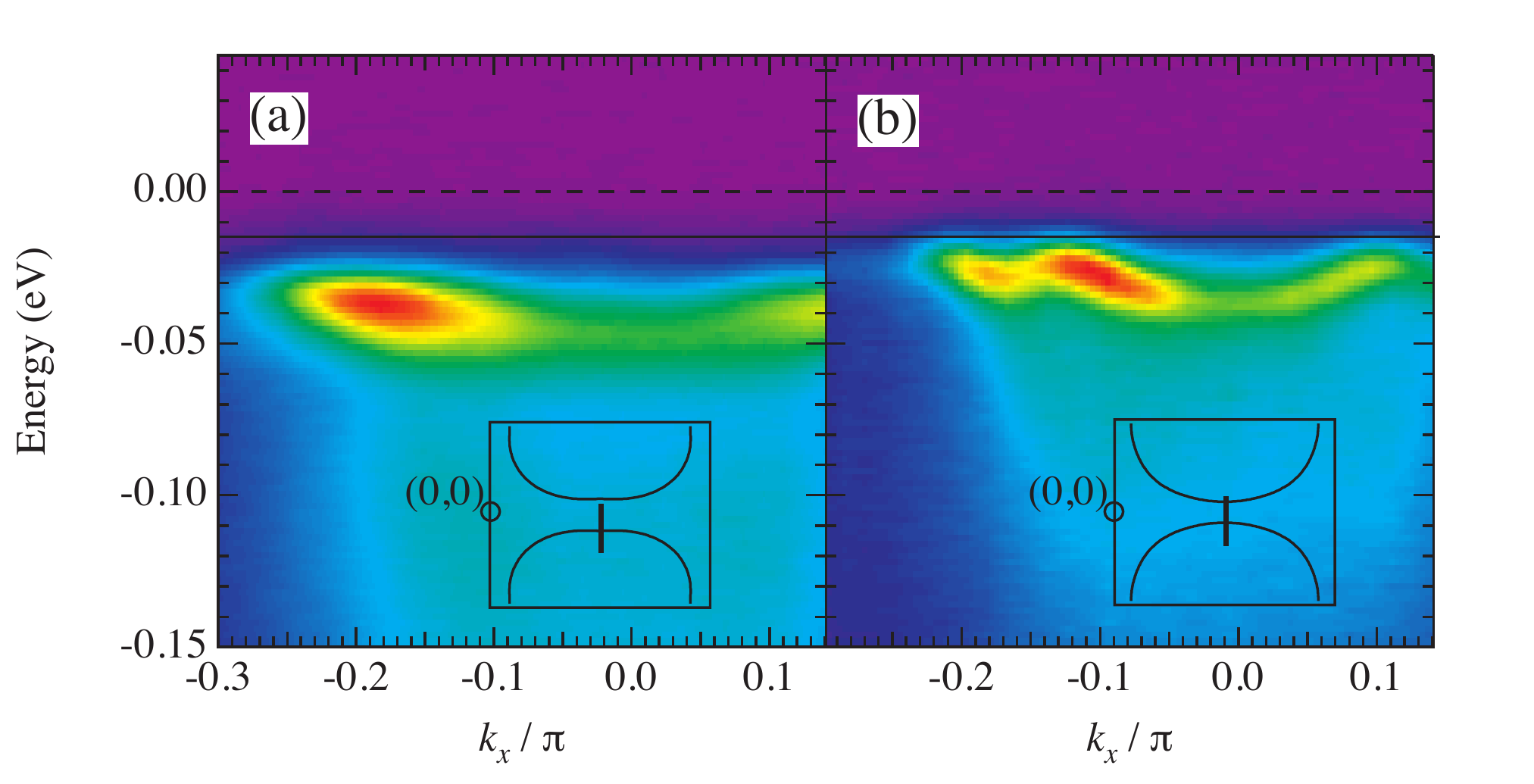}
\caption{(Color online) (a) ARPES intensity map of freshly cleaved optimally doped Bi2212 at ($\pi$, 0) showing no bilayer splitting, (b) ARPES intensity maps on the same sample and the same location as in (a) only oxygen aged (\textit{in-situ} overdoing) in a UHV system with a leak showing bilayer splitting and a peak shift location of the Fermi momentum, with the black line as a guide to the eye.}
\label{Fig. 2}
\end{figure}

n absence of leaks, a reasonable UHV system has normally undetectable levels of oxygen. However in stainless steel vessels CO and CO$_2$ are always present. These oxide molecules can adhere to clean sample surfaces especially at low temperatures. When the molecules are exposed to VUV photons above 6 eV they break into carbon and oxygen \cite{M. M. Halmann}; the oxygen can then be incorporated into BiO layer as dopant, while the carbon atoms remain on the surface. The proof of this scenario is in Fig. 1 (e) where the core-level spectrum of Bi2212 at 300K and 40K are shown.  As the sample cooled more CO and CO$_2$ molecules adhered to the surface of the sample. Since there are carbon deposits some time after cleaving and even more after cooling, it is likely the oxygen accompanied the carbon to the surface.  This oxygen can then change the doping of the sample after it is dissociated from the carbon. 

\begin{figure}
\includegraphics[width=3.2in]{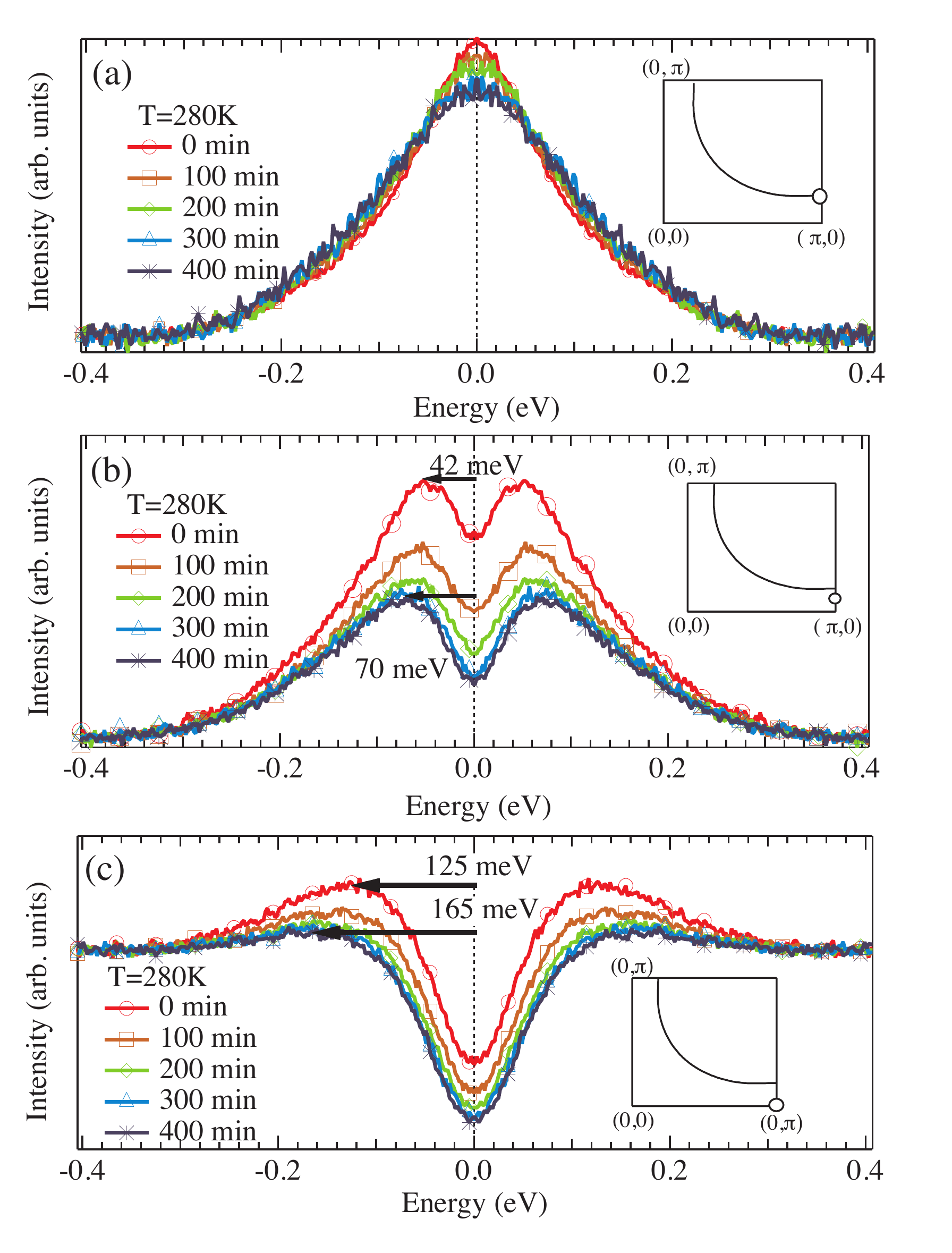}
\caption{(Color online) (a)-(c) symmetrized ARPES EDC's for Bi2212 taken at three points near ($\pi$,0) showing the time evolution of the spectrum at 280K.}
\label{Fig. 2}
\end{figure}

In the presence of a leak a UHV system can have detectable amounts of oxygen.  Under these conditions a Bi2212 sample can age even without the breakdown of CO and CO$_2$.  One of the trademarks of an over-doped (aged) Bi2212 sample is the appearance of bi-layer band splitting at the antinode ($\pi$, 0).  While there has been a relatively active discussion on whether Bi2212 contains bilayer band splitting all the time or just in an over-doped state; bilayer splitting has only been seen in over-doped samples when using a helium discharge lamp \cite{Y.-D. Chaung 2004, S. V. Borisenko 2004, S. V. Borisenko 2006, A. A. Kordyuk 2004}. An example of this is shown in FIG. 2 where a fresh Bi2212 sample was scanned and then allowed to sit in the leaky UHV system overnight before scanning again. Even though the sample was kept a 20 K, bilayer band splitting was detected after the break, signaling that the sample aged because of oxygen absorption.

\section{Decreasing carrier concentration}
\begin{figure}
\includegraphics[width=3.6in]{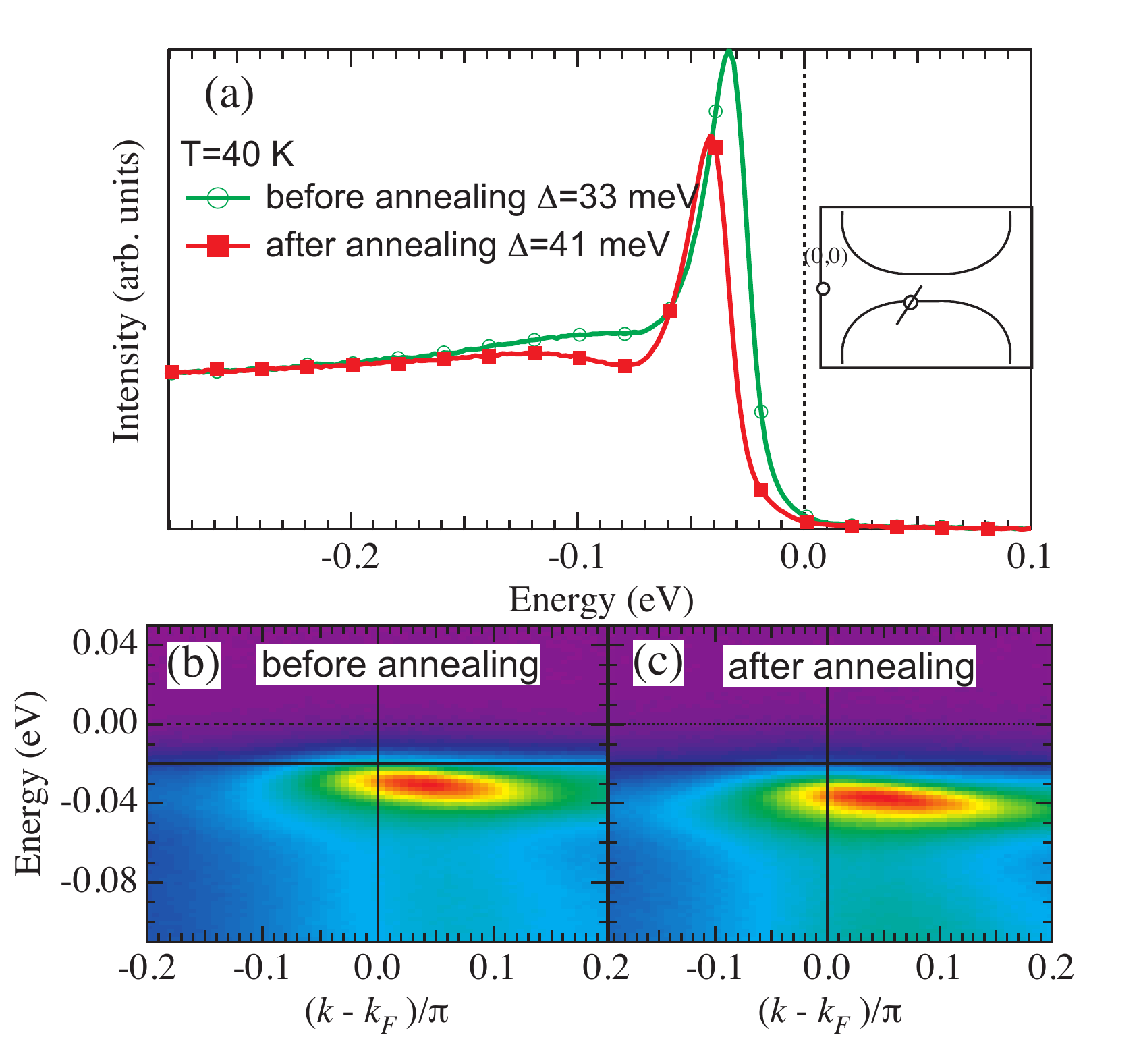}
\caption{(Color online) (a) EDC at the Fermi momentum close to the anti-node before (green circles) and after (solid red squares) annealing at 280K over 28 hours with their respective superconducting gaps $\Delta$, (b)-(c) momentum intensities maps taken across Fermi momentum close to ($\pi$, 0) before and after annealing.}
\label{Fig. 3}
\end{figure}

While in a reasonable vacuum system there can be enough CO$_2$/CO to change the surface doping of a sample over time; in an ultra clean UHV system samples can live for many weeks without surface degradation or a change in doping (assuming the sample is kept at low temperature).  Yet, when the sample is annealed above 200K an interesting thing happens to the Bi2212's doping level; the sample doping level is reduced (the opposite of aging).  This is seen in Fig. 3 (a)-(c) where the time evolution of Bi2212's EDCs at three locations at or near ($\pi$,0) with the sample at 280K is shown.  The sample actually changes doping moving towards lower doping (signified by a larger spectral gap). Fig. 4 (a) shows the energy distribution curve (EDC) at the anti-nodal Fermi momentum from the same sample before and after annealing at 280K for 28 hours.  The superconducting gap clearly shifts from 33 meV to 41 meV and the peak is suppressed, signaling that the doping has changed from a slightly over doped sample to a more under doped sample\cite{T. Sato 2001}.   The momentum color maps from Fig. 4 (a) are shown in FIG. 4 (b)-(c); after annealing the gap shifts to higher binding energy, there is also a shift in the location of the Fermi momentum.  This momentum shift comes from a change in the chemical potential, which moves lower in a ridged-band-like fashion upon doping.\cite{M. Hashimoto 2008}

Another way to see if a samples carrier concentration has decreased is to look at the pseudogap.  Fig. 5 (a) shows the EDC at the Fermi momentum before and after annealing at 280K for 28 hours.  The pseudogap shifts from 30 meV to 50 meV.  As Bi2212 goes to lower doping levels the pseudogap becomes bigger and the temperature at which the pseudogap remains (T*) becomes higher\cite{H. Ding 1996}. Fig. 5 (b)-(c) demonstrates that before annealing T* is below 140K with the pseudogap disappearing and after annealing T* is above 200K.  The pseudogap after annealing is above 200K, which guarantees that the sample is at a lower doping level.

\begin{figure}
\includegraphics[width=3.6in]{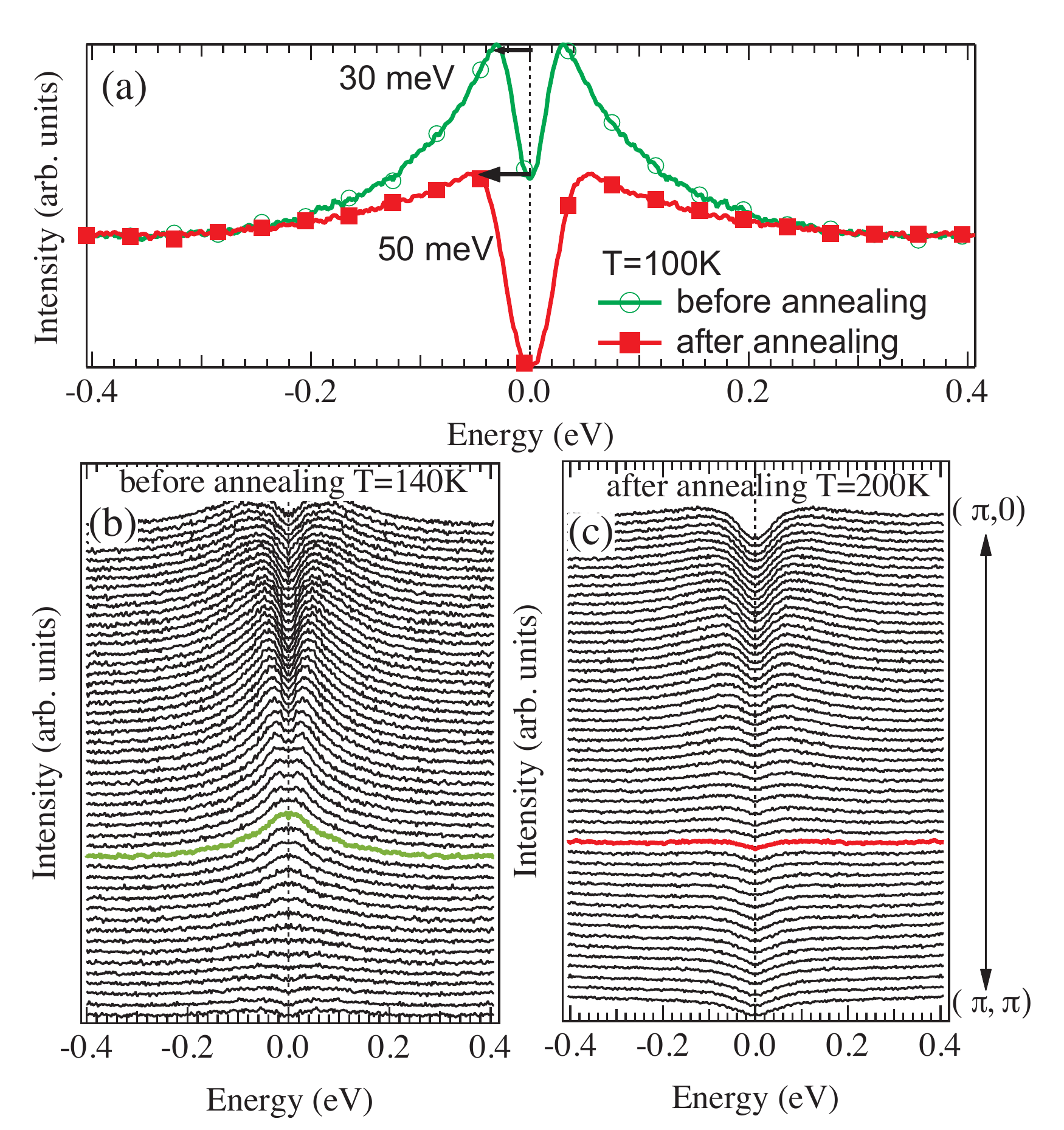}
\caption{(Color online)  (a) 100 K symmetrized ARPES data taken at the Fermi momentum before and after annealing at 280 K for 28 hours, (b) ARPES intensities at 140 K before annealing, (c) ARPES intensities at 200 K after annealing.}
\label{Fig. 4}
\end{figure}

Until now we have only shown the lowering of doping on Bi2212 at elevated temperature.  While we still haven't shown if the doping change is caused by the elevated temperature or a combination of elevated temperature and VUV photons.  This was tested by scanning the sample just after cleaving and again after the sample sat under UHV for 16 days at 100K. This data is shown in figure Fig. 6 (a).  The spectrum barely changed over the two weeks. While in Fig. 6 (b) we show the 280K spectrum just after cleaving, and again after the sample sat under UHV for 8 days at 280K.  Most of the spectral weight has shifted to higher energies and the Fermi edge has all but disappeared, signifying an almost completely insulating sample. From Fig. 6 we can conclude that the lowering of the samples doping is only caused by the elevated temperatures.

\begin{figure}
\includegraphics[width=3.6in]{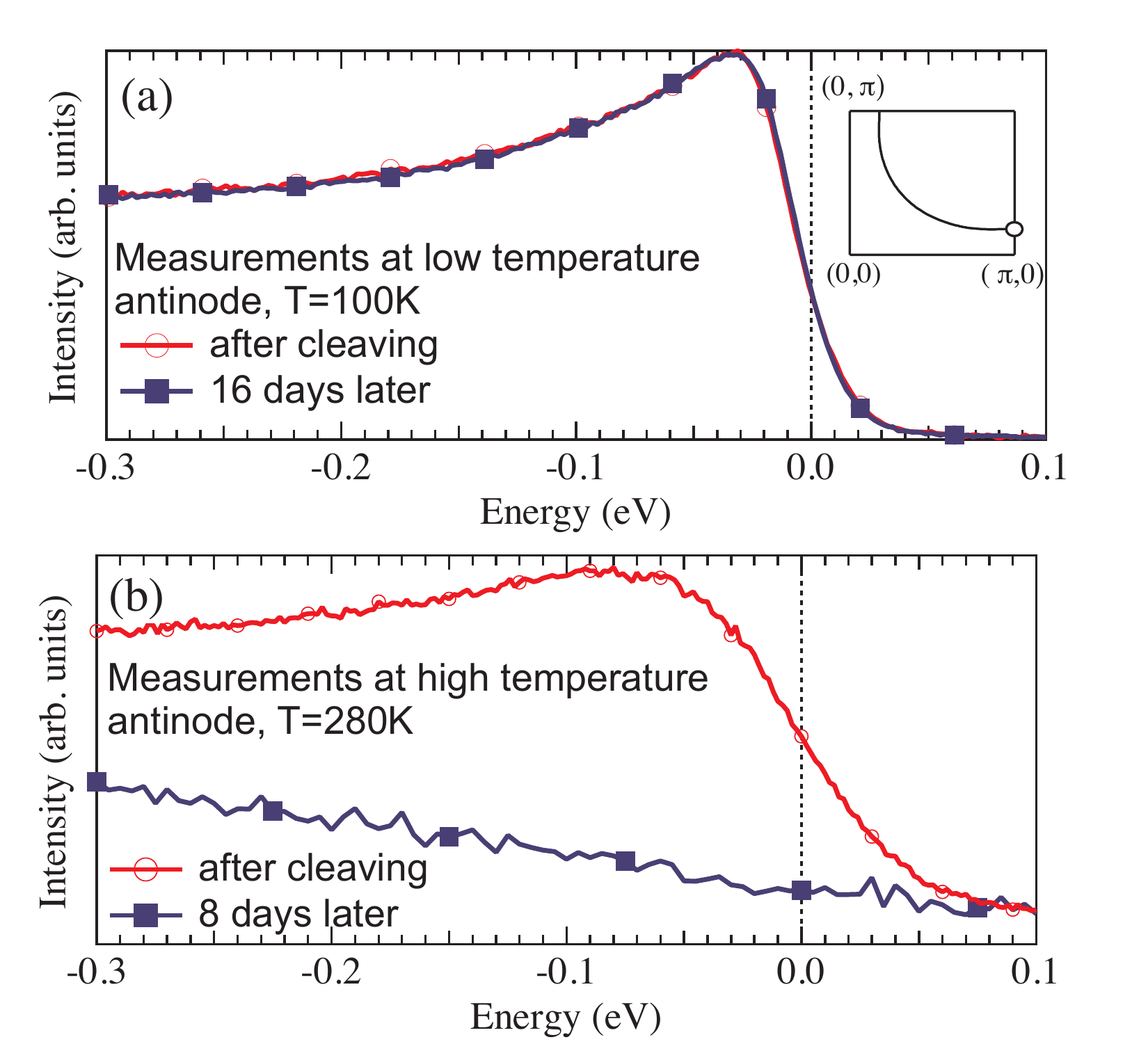}
\caption{(Color online) Bi2212 EDC at the Fermi momentum close to ($\pi$,0) (a) just after cleaving at 100K (red circles) and again after sitting at 100K for 16 days (solid blue squares), (b) just after cleaving at 280K (red circles) and again after sitting at 280K for 8 days (solid blue squares).}
\label{Fig. 5}
\end{figure}

\begin{figure*}
\includegraphics[width=6 in]{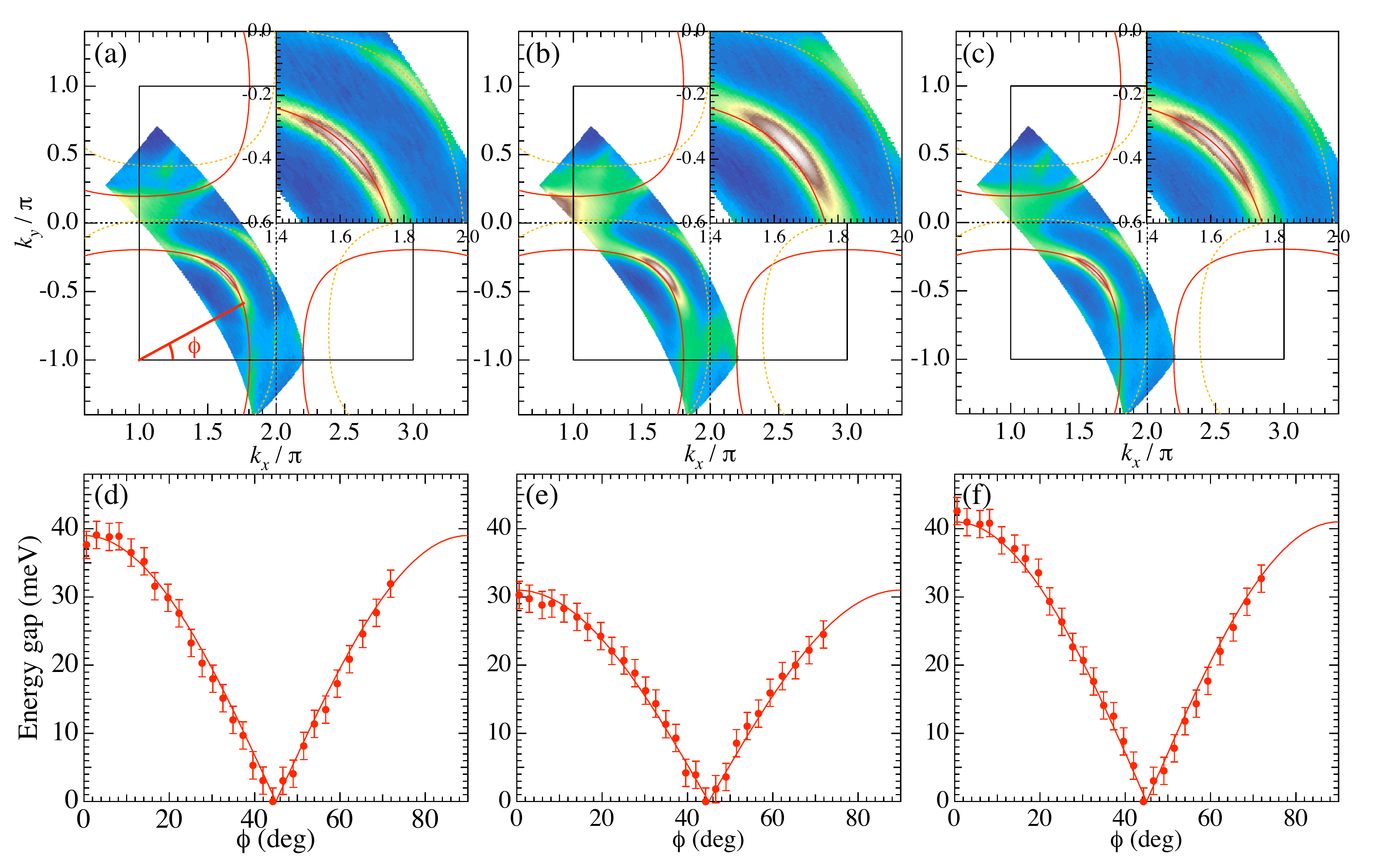}
\caption{(Color online) ARPES intensity plots at the Fermi energy from the same sample at three different times all at 12K: (a) just after cleaving, (b) after a couple of days of VUV aging at low temperature, (c) after annealing at 280K overnight; the upper right hand corner of (a)-(c) are the zoomed in images from the bottom left hand corner, the red and dotted yellow curves are from a tight binding fit for optimally doped Bi2212 as a guide to the eye, (d)-(f) the size of the superconducting gap as a function of angle $\phi$ from (a)-(c) respectively.}
\label{Fig. 6}
\end{figure*}

The greatest consequence of this study is that Bi2212's doping can be change from over doped all the way down to insulating in a systematic fashion on a single crystal. To this point the data presented has been either over doped by aging or under doped by annealing on different samples.  Fig. 7 demonstrates how a \textit{single} sample can be over-doped by aging and then under-doped by annealing to move across the phase diagram. An optimally doped Bi2212 sample was cleaved, the Fermi surface and superconducting gap values as a function of angle $\phi$ (angle clockwise from the line ($\pi$,-$\pi$) to (2 $\pi$,-$\pi$)) was scanned Fig. 7 (a) $\&$ (d).  Aging was detected after a couple of days of scanning Fig. 7 (b) $\&$ (e).  The sample was then annealed overnight at 280K to remove the aging Fig. 7 (c) $\&$ (f).

\section{Conclusion}

We have presented a systematic study of the electronic properties at the surface of Bi2212 as a function of vacuum conditions.  The results confirm that under poor vacuum conditions there is an increase in carrier concentration due to the breakup of CO and CO$_2$ molecules by exposure to vacuum ultra-violet (VUV) photons and a subsequent adsorption of oxygen into the BiO layers. We also show that with a UHV leak a sample can increase its carrier concentration just by sitting in the vaccum.  This observation confirms that bilayer splitting only occurs in over-doped Bi2212. We then show that at elevated temperatures (T$>$200K) the sample surface loses oxygen, which results in a reduction of the carrier concentration.  These two effects (\textit{in-situ} absorption and desorption of oxygen) can be utilized in order to control the carrier concentration of Bi2212. This approach enables one to study the intrinsic electronic properties (i.e. without changing the impurities and defects) of the cuprates across the phase diagram in ARPES as well as other surface sensitive techniques on a single sample.

\section{Acknowledgments}

This work was supported by Director Office for Basic Energy Sciences, US DOE. Work at Ames Laboratory was supported by the Department of Energy - Basic Energy Sciences under Contract No. DE-AC02-07CH11358.  The work at BNL was supported by Department of Energy - Basic Energy Sciences under Contract No. DE-AC02-98CH10886. Synchrotron Radiation Center is supported by the National Science Foundation under award No. DMR-0537588.

\end{document}